\documentclass[prl,twocolumn,showpacs,superscriptaddress,10pt]{revtex4-1} 
\usepackage{amsmath,amssymb,graphicx,bm}

\usepackage{color,times}
\usepackage{sidecap}

\definecolor{darkblue}{rgb}{0, 0, 0.8}
\usepackage[colorlinks=true, breaklinks=true, linkcolor=darkblue, citecolor=darkblue, urlcolor=darkblue]{hyperref}
\newcommand{\ket}[1]{\left| #1\right\rangle}
\newcommand{\bra}[1]{\left\langle #1\right|}

\definecolor{rkcol}{rgb}{0.8, 0, 0.5}

\begin{document}
	
	\title{Accurate mapping of multilevel Rydberg atoms on interacting spin-$1/2$ particles \\for the quantum simulation of Ising models}
	
	\author{Sylvain de L\'es\'eleuc$^\ast$}
	\affiliation{Laboratoire Charles Fabry, Institut d'Optique Graduate School, CNRS, Universit\'e Paris-Saclay, 91127 Palaiseau Cedex, France}
	\author{Sebastian Weber$^\ast$}
	\affiliation{Institute for Theoretical Physics III and Center for Integrated Quantum Science and Technology, University of Stuttgart, 70550 Stuttgart, Germany}
	\author{Vincent Lienhard}
	\author{Daniel~Barredo}
	\affiliation{Laboratoire Charles Fabry, Institut d'Optique Graduate School, CNRS, Universit\'e Paris-Saclay, 91127 Palaiseau Cedex, France}
	\author{Hans Peter B\"uchler}
	\affiliation{Institute for Theoretical Physics III and Center for Integrated Quantum Science and Technology, University of Stuttgart, 70550 Stuttgart, Germany}
	\author{Thierry Lahaye}
	\author{Antoine Browaeys}
	\affiliation{Laboratoire Charles Fabry, Institut d'Optique Graduate School, CNRS, Universit\'e Paris-Saclay, 91127 Palaiseau Cedex, France}
	
\begin{abstract}
We study a system of atoms that are laser-driven to $nD_{3/2}$ Rydberg states and assess how accurately they can be mapped onto spin-$1/2$ particles for the quantum simulation of anisotropic Ising magnets. Using non-perturbative calculations of the pair interaction potentials between two atoms in the presence of both electric and magnetic fields, we emphasize the importance of a careful selection of the experimental parameters in order to maintain the Rydberg blockade and avoid excitation of unwanted Rydberg states. We then benchmark these theoretical observations against experiments using two atoms. Finally, we show that in these conditions, the experimental dynamics observed after a quench is in good agreement with numerical simulations of spin-1/2 Ising models in systems with up to 49 spins, for which direct numerical simulations become intractable. 
\end{abstract}
	
\maketitle 

A promising approach for quantum information science and for quantum simulation relies on single atoms trapped in optical tweezers and excited to Rydberg states~\cite{Saffman2010}. Recent experimental progress has demonstrated the active loading of up to $ 50$ atoms in arrays of optical tweezers arranged in arbitrary geometries with a controllable separation between the atoms~\cite{Barredo2016,Endres2016}. The strong interactions between the Rydberg atoms (van der Waals or dipolar exchange) make these systems ideal for quantum simulation of, e.g., spin Hamiltonians~\cite{Saffman2010,Glaetzle2015}, lattice gauge theories \cite{Glaetzle2014}, or systems characterized by topological invariants~\cite{Gorshkov2013,Peter2015}.

\begin{figure}[t]
\centering
	\includegraphics[scale=1.0]{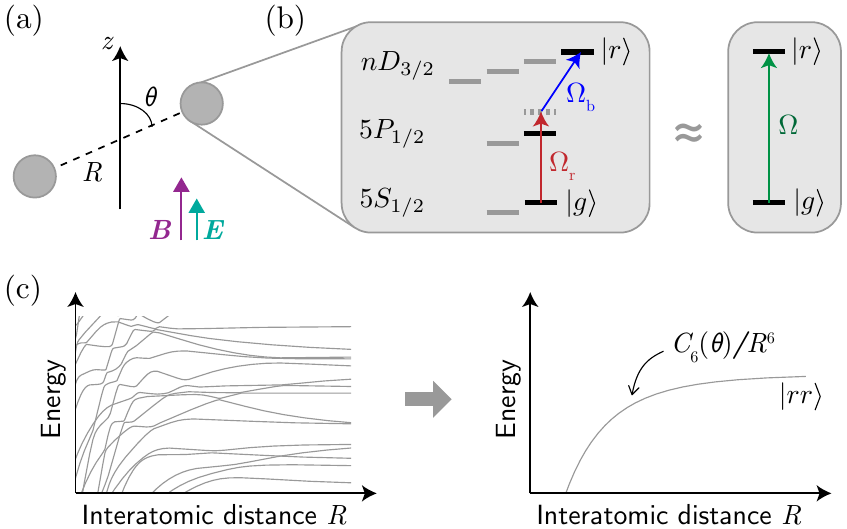}
	\caption{Mapping a system of multilevel Rydberg atoms onto a spin-$1/2$ model. 
	(a) System: two atoms separated by a distance $R$; $\theta$ is the angle between the interatomic axis and the quantization axis $z$ defined by a magnetic field $B$. An electric field $E$ can be applied along $z$. (b) A two-photon transition couples coherently the ground state $\ket{g}$ to a target Rydberg state $\ket{r}$ with an effective two-photon Rabi frequency $\Omega$. (c) Full energy spectrum of the atom pair. The mapping consists in replacing this complex structure by an effective interaction potential.}
	\label{fig:fig1}
\end{figure}

One of the main ingredients to realize such pristine artificial systems is the identification of suitable Rydberg levels and a full characterization of the interaction potentials. In the simplest case one identifies the atomic ground state as the spin-down state $\ket{\downarrow}$ and the Rydberg excitation as the spin-up state $\ket{\uparrow}$ for the implementation of spin-1/2 Hamiltonians~\cite{Schauss2012,Schauss2015,Labuhn2016,Bernien2017}. However, in practice, describing the atom as a two-level system is an approximation that can be difficult to fulfill due to the small splittings between levels in the Rydberg manifold. For a single atom, it is sufficient to apply a magnetic field of a few Gauss in order to isolate a single two-level transition. But already for two atoms, the density of pair-states becomes quite large, and, due to the interactions, mixing between different levels occurs in configurations without special symmetries (Fig.~\ref{fig:fig1}). Finding the optimal parameters such that the system is accurately described as a spin-1/2 system with a well-defined interaction potential is thus a non-trivial task, that needs to be addressed in view of applications in quantum simulation. 
	
A natural choice for implementing spin Hamiltonians with Rb Rydberg atoms is to use $nS$ Rydberg states~\cite{Bernien2017}, as they possess only two Zeeman sublevels and do not feature F\"orster resonances~\cite{Reinhard2007}. However, many experiments use $nP$ or $nD$ states: the former are the only ones accessible from the ground state using single-photon dipole transitions~\cite{Tong2004,Hankin2014} and are used in particular for Rydberg dressing~\cite{Zeiher2016,Biedermann2016,Lee2017,Zeiher2017}, while the latter~\cite{Urban2009,Gaetan2009,Labuhn2016} require less laser power for excitation from the ground state as compared to $nS$ states. Moreover, for both $nP$ and $nD$ states, the van der Waals interaction can be anisotropic, opening the way for simulating exotic matter~\cite{Glaetzle2014,Glaetzle2015}. Nevertheless when implementing an anisotropic Ising model with $nD_{3/2}$ states, deviations from the prediction of a spin-1/2 model can occur, as we observed in Ref.~\cite{Labuhn2016}. 

In this Letter, we thus focus on Rydberg $nD_{3/2}$ states, and derive under which conditions the simple picture of a spin-1/2 model with an effective anisotropic interaction potential between the pair states is valid, despite the large number of Rydberg levels involved. For that purpose, we use recently developed open-source software~\cite{Weber2017, PairinteractionHomepage} to numerically calculate the exact pair-state potentials in the presence of external electric and magnetic fields. We find a remarkable sensitivity of the interaction spectrum to weak static electric fields, which can lead to a breakdown of the Rydberg blockade not considered in previous studies~\cite{Walker2005,Reinhard2007,Walker2008,Pohl2009,Vermersch2015,Derevianko2015}. We then experimentally corroborate this prediction in a simple system of two atoms. Finally, we extend our study to a ring of 8 atoms and a $7\times7$ square array, the settings for which some deviations from the spin-1/2 model were observed in~\cite{Labuhn2016}, and now demonstrate a much better agreement with a numerical simulation. 
	
We use the Rydberg state $\ket{r}=\ket{nD_{3/2},m_J=3/2}$ and couple it to the atomic ground state $\ket{g}=\ket{5S_{1/2},F=2,m_F=2}$ by a  two-photon transition, see Fig.~\ref{fig:fig1}(b). Ideally we want to identify the states  $\ket{g}$ and $\ket{r}$ with pseudo spin-$1/2$ states $\ket{\downarrow}$ and $\ket{\uparrow}$. In this case, when taking into account interactions between atoms in $\ket{r}$, the system maps onto an Ising-like model in a transverse field~\cite{Schauss2012,Schauss2015,Labuhn2016,Bernien2017} governed by the Hamiltonian
\begin{equation}
	H=\sum_i\frac{\hbar\Omega}{2}\sigma_x^i+\frac{1}{2}\sum_{i \neq j} U_{ij} n_{i} n_{j}.
	\label{eq:spin12}
\end{equation}
Here, $\Omega$ is the Rabi frequency corresponding to the laser driving, $\sigma_x^i=\ket{r}\bra{g}_i+\ket{g}\bra{r}_i$ and $n_i=\ket{r}\bra{r}_i$, and the rotating wave approximation has been applied. The interaction between atoms $i$ and $j$ is given at large distances by an anisotropic van der Waals potential $U_{ij}=C_6(\theta_{ij})/R_{ij}^6$, where $R_{ij}$ is the interatomic distance and $\theta_{ij}$  the angle between the internuclear axis and the quantization axis, see Fig.~\ref{fig:fig1}(a). For shorter distances, deviations from the $1/R^6$ behavior are expected (see below).
		
\begin{figure}[t!]
	\centering
	\includegraphics[scale=1.0]{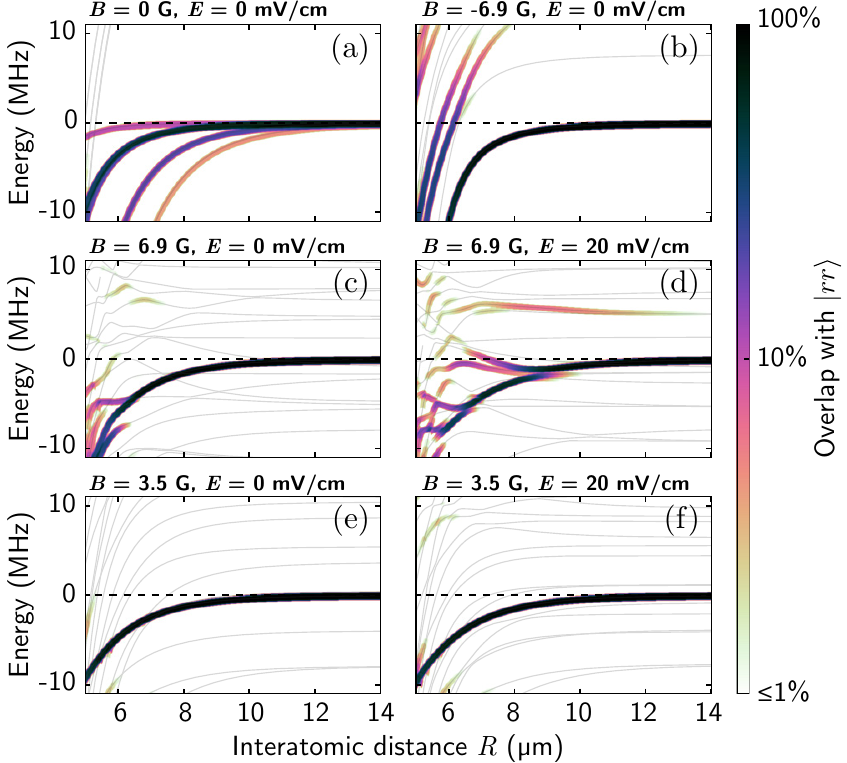}
	\caption{Influence of magnetic and electric fields on the interaction potentials around the pair-state $\ket{rr}$ where $\ket{r} = \ket{61D_{3/2},m_j = 3/2}$, for $\theta=78^\circ$. The shading encodes the overlap of the eigenstates with the non-interacting state $\ket{rr}$. (a) $B=0$ and $E=0$: $\ket{rr}$ overlaps with all the degenerate Zeeman pair states. (b) $B=-6.9$~G and $E=0$: the interaction curves are split due to the Zeeman effect. Some curves still strongly mix with $\ket{rr}$ due to the interaction. (c) $B=6.9$~G and $E=0$: one potential curve dominates. However, (d) the addition of a small electric field $E=20$~mV/cm is enough to strongly perturb the pair states. (e-f) This behavior is absent for $B=3.5$~G.}
	\label{fig:fig2}
\end{figure}

We now look for conditions allowing us to describe the interaction spectrum for a pair of atoms by a single potential curve $U(R,\theta)$ as shown in Fig~\ref{fig:fig1}(c). To approach this problem quantitatively, purely analytic approaches are of little use, and we use numerical methods to diagonalize  the dipole-dipole Hamiltonian~\cite{Sibalic2016} (as well as higher-order multipole contributions) in the presence of arbitrarily oriented external electric and magnetic fields~\cite{Weber2017}. In view of reproducing the experiment of Ref.~\cite{Labuhn2016}, we chose the state $\ket{r}=\ket{61D_{3/2},m_J=3/2}$. Figure~\ref{fig:fig2} shows the interaction spectrum for a generic angle $\theta=78^\circ$. The shading of the various interaction potentials shows the overlap of the states with $\ket{rr}$. In panel (a), no magnetic and electric fields are applied, and some Zeeman pair states interact very weakly, while they are still coupled to $\ket{gg}$. Consequently, the Rydberg blockade is broken as the double excitation of Rydberg states is possible even at short distances~\cite{Walker2005,Walker2008}. Panel~(b) shows the interaction potentials, but now in the presence of a magnetic field $B= -6.9$~G. The Zeeman effect splits the various potentials and the state $\ket{rr}$ is now well isolated from the other eigenstates. However, since the sign of the Zeeman shift is opposite to that of the van der Waals interaction, there are some specific values of the interatomic distance $R$ where the laser excitation of other Zeeman pair states is resonant; these `magic distances', predicted by \cite{Vermersch2015}, can thus lead to a breakdown of the blockade. In order to avoid this effect, one can simply use an opposite value for the $B$ field, as in panel (c), where $B=6.9$~G. These parameters are similar to the ones used in~\cite{Labuhn2016}, and in these conditions, it is a good approximation to describe the system by a single state for $R>6 \, \mu$m.

\begin{figure}[t]
	\centering
	\includegraphics[scale=1.0]{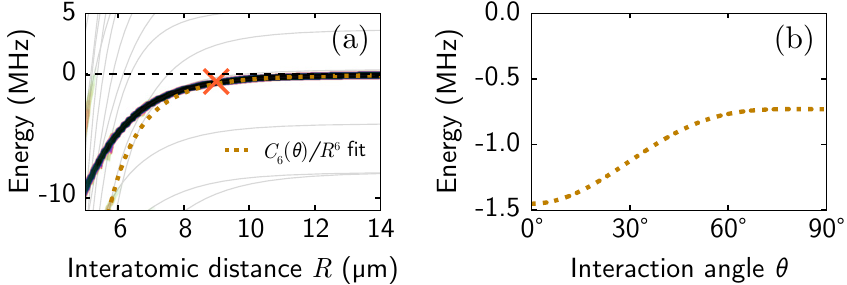}
	\caption{Approximation of the interaction by an anisotropic van der Waals potential $C_6(\theta)/R^6$. (a) Comparison of the exact interaction energy (solid line) with the asymptotic determination of the van der Waals potential (dashed line) for a fixed angle $\theta=78^\circ$ and $B = 3.5$~G. (b) Angular dependence of $C_6(\theta)/R^6$ at $R=9\, \mu$m marked by the cross on (a). 
	}
	\label{fig:fig6}
\end{figure}

It turns out however that the interaction potentials are extremely sensitive to electric fields $E$. Figure~\ref{fig:fig2}(d) corresponds to the same parameters as in panel (c), but now in the presence of an electric field $E=20$~mV/cm along $z$. A naive calculation of the Stark shift of pair states for such a value of $E$, neglecting resonance effects, would give shifts in the 100~kHz range, which would have hardly any influence on the potentials. However, the exact diagonalization shows that the interaction potentials are strongly affected, with many states being resonant with the  excitation laser. We thus expect a significant breakdown of the Rydberg blockade in these conditions. Remarkably, this effect is absent for lower magnetic fields $B=3.5$~G, see Fig.~\ref{fig:fig2}(e-f). In the optimal regime where a single potential curve $U(R,\theta)$ can be identified, we check if we can describe it by a van der Waals potential with an angular dependence $C_6(\theta)/R^6$. Figure \ref{fig:fig6}(a) shows the energy dependence as a function of $R$ for $\theta = 78^\circ$ together with a $1/R^6$ fit. We observe, that for $R \gtrsim 8 \, \mu$m, the van der Waals description is an excellent approximation. Figure \ref{fig:fig6}(b) shows the angular dependence of the coefficient $C_6(\theta)$. We have thus extended the anisotropic effective potential approach developed in ~\cite{Barredo2014,Vermersch2015} beyond the strong blockade regime.

We now turn to the experimental test of the above predictions. Our setup has been described in detail elsewhere \cite{Labuhn2016}: we create two-dimensional arrays of optical tweezers ($1/e^2$ radius of $1 \, \mu$m, depth of $1$~mK), in which we load single atoms from a magneto-optical trap. Active sorting of the atoms with a moving tweezers allows us to obtain fully-loaded arrays~\cite{Barredo2016} with up to 49 atoms. We optically pump the atoms into $\ket{g}$ in the presence of a magnetic field pointing along the $z$ axis, within the arrays' plane. We then switch off the tweezers and illuminate the atoms with a Rydberg excitation pulse of duration $\tau$ (we use a two-photon transition with lasers at 795 and 475~nm detuned from the $5P_{1/2}$ intermediate state giving an effective Rabi frequency $\Omega = 2\pi \times 1.2$~MHz). At the end of the sequence, we switch on again the tweezers. Atoms in $\ket{g}$ are recaptured in the tweezers while those that have been excited to Rydberg states (either in $\ket{r}$ or in any other Rydberg state) are repelled by the traps and lost from the trapping region~\cite{Labuhn2016,Leseleuc_rabi}. Thus, when we switch on again the MOT beams, atoms in $\ket{g}$ are observed by fluorescence, while missing atoms are assigned to Rydberg states.

\begin{figure}[t]
	\centering
	\includegraphics[scale=1.0]{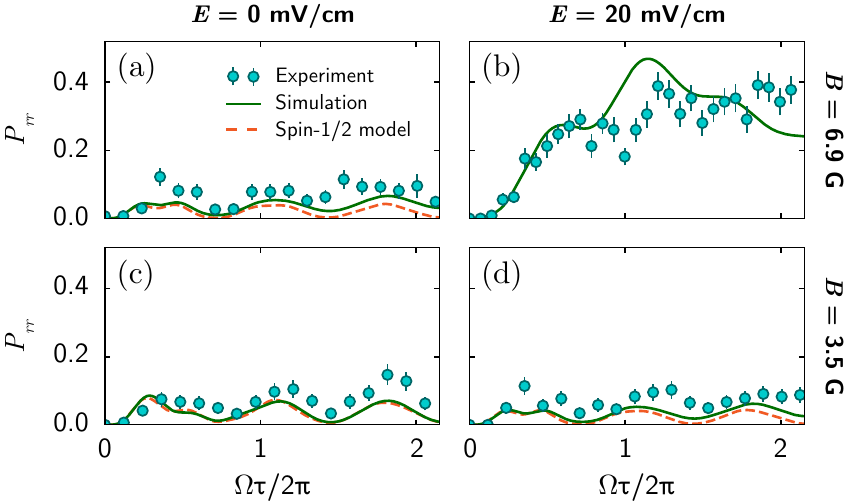}
	\caption{Two-atom blockade experiments. Probability $P_{rr}$ to excite the two atoms as a function of the pulse area $\Omega \tau$. For $B=6.9$~G (a-b), increasing $E$ from 0 to 20~mV/cm breaks the Rydberg blockade. At $B=3.5$~G (c-d), an efficient blockade is maintained, even in the presence of the electric field. The solid lines result from a simulation taking into account the full interaction spectrum (see text). The dashed lines are obtained by modeling the atoms as spin-1/2 particles with a single interaction potential for $\ket{rr}$, except in case (b) where the pair-state is too perturbed. The error bars show the standard error of the mean.}
	\label{fig:fig4}
\end{figure}

As a first test of the influence of electric and magnetic fields on the potential curve, we perform two-atom blockade experiments~\cite{Urban2009,Gaetan2009} with $R= 6.5 \, \mu$m and $\theta=78^\circ$, i.e., the same parameters as in Fig.~\ref{fig:fig2}.  We use four different settings of the external fields: the magnetic field is either 3.5 or 6.9~G, and the electric field either zero (within the accuracy $\sim 5$~mV/cm of our cancellation of stray fields) or $20$~mV/cm. In order to quantify the Rydberg blockade, we measure the probability $P_{rr}$ to have two Rydberg excitations after illuminating the atoms with the excitation pulse. The results are displayed in Fig.~\ref{fig:fig4}. We observe as expected a strong suppression of $\ket{rr}$  for all settings, except for $B= 6.9$~G and $E=20$~mV/cm, where we find a significant probability to excite the two atoms. To compare with the theory, we simulate the dynamics of the two-atom system solving the Schr\"odinger equation and calculate the probability to excite the two atoms~\cite{footnote_fluctuation}. We assume two different models to describe the interacting system: in the first one (Fig.~\ref{fig:fig4} solid line), we use the full interaction spectrum and include around 800~pair-states within 2~GHz from the resonance (a bigger electric field would drastically increase the basis size). In the second model (dashed line), we describe the interaction in the $\ket{rr}$ state with the single potential curve identified above, thus solving the spin-1/2 model governed by the Hamiltonian~(\ref{eq:spin12}). This simulation with no adjustable parameter is in excellent agreement with the experimental data.

\begin{figure}[b]
	\centering
	\includegraphics[scale=1.0]{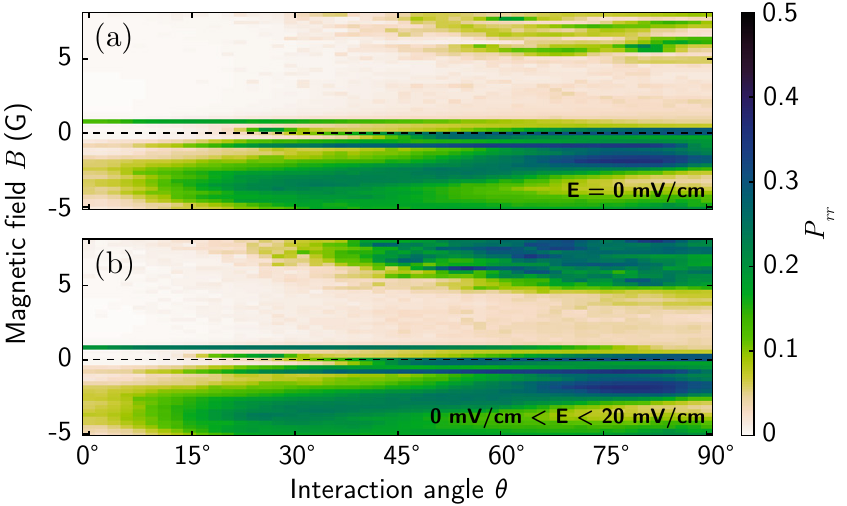}
	\caption{ 
		Influence of $\theta, B, E$ on the mapping onto a spin-$1/2$ system. Probability of double excitations at long times (see text) as a function of the magnetic field $B$ and the angle $\theta$. The interatomic distance is fixed at $R=6.1 \, \mu$m. The electric field is $E = 0$ in (a) and chosen between $0$ and $20$~mV/cm such that the probability for two Rydberg excitations is maximized in (b).}
	\label{fig:fig3}
\end{figure}

\begin{SCfigure*}
	\centering
	
	\includegraphics[scale = 1]{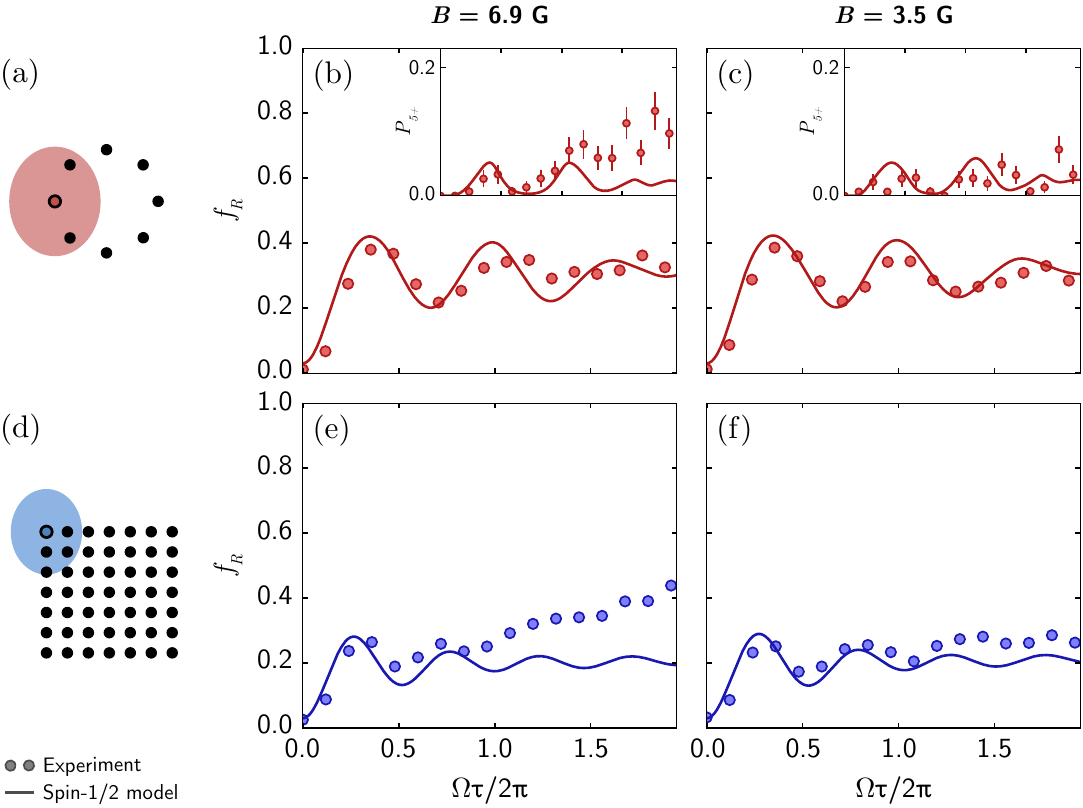}
	\hspace*{5mm}
	\caption{Dynamics of an ensemble of atoms under Rydberg excitation.	(a) 8-atom ring with a nearest neighbor spacing of $6.5 \, \mu$m. The shaded ellipse illustrates the range of the anisotropic blockaded region $U>\hbar \Omega$. (b) Evolution of the Rydberg fraction $f_{\rm R}$ with the pulse area $\Omega \tau$ for $B=6.9$~G. The inset shows the probability $P_{5+}$ to observe configurations with at least 5 excitations. At large times, the experimental points systematically lie above the results of a simulation of the corresponding spin-$1/2$ model (solid line). (c) Same parameters with $B=3.5$~G. (d) Square lattice of $7\times 7 $ traps (lattice spacing $6.1 \, \mu$m). The blockade extends over nearest and next-nearest neighbors. (e) Evolution of the Rydberg fraction for $B=6.9$~G. Here the data shows a slow increase in $f_{\rm R}$ at long times, while the spin-$1/2$ model predicts a saturation. (f) For $B=3.5$~G, the agreement with the spin-$1/2$ model becomes very good. All figures: error bars depict the standard error of the mean and are often smaller than the symbol size.}
	\label{fig:fig5}
	
\end{SCfigure*}

We now investigate more systematically how the geometry and the value of the electric and magnetic fields affect the accuracy of the mapping on a spin-$1/2$ model. Using the exact simulation taking into account the full interaction spectrum, as done for Fig.~\ref{fig:fig4}, we calculate the average value of the double excitation probability $P_{rr}$ at long times and look at the range of parameters for which $P_{rr}$ remains small. Panel (a) corresponds to the case $E=0$, while panel (b) shows a `worst-case scenario' where $E$ is chosen in the range $0-20$~mV/cm so as to maximize $P_{rr}$. For $\theta \approx 0$ the system is faithfully described by a spin-1/2 system. For increasing angle $\theta$, we identify the range of magnetic fields where Rydberg blockade is maintained. In addition, we observe a breaking of the Rydberg blockade for negative magnetic fields as predicted in~\cite{Vermersch2015}. A similar analysis for various principal quantum numbers $n$ indicates that the presence of a F\"orster resonance at $n=59$ is responsible for this sensitivity to weak electric fields~\cite{Ravets2014a}.

Now that we have identified parameters allowing to map our two-atom system onto a spin-1/2 model, we extend the study to larger systems. We first revisit the experimental realization of an 8-atom ring, reported in Ref.~\cite{Labuhn2016}, where we observed a discrepancy with the spin-$1/2$ model. We illuminate the atoms with a Rydberg excitation pulse and observe the ensuing dynamics following this quench by measuring the fraction $f_{\rm R}$ of atoms that are excited to Rydberg states. We also extract the probability $P_{5+}$ that more than five atoms are excited, i.e., that the blockade condition is violated as for our parameters nearest-neighbor excitation is thwarted. Prior to this experiment we compensated the stray electric field better than $5$~mV/cm. Figure~\ref{fig:fig5}(a-c) shows the results for two values of the magnetic fields. For $B=6.9$~G, we observe a slow rise of $P_{5+}$ above the prediction of the spin-$1/2$ model. Contrarily, for $B=3.5$~G, we find a much better agreement with the spin-1/2 model as expected from the above analysis.
		
In a second experiment, we probe a square array of $7\times7$ atoms. The evolution of $f_{\rm R}$ is shown in Fig.~\ref{fig:fig5}(d-f). As an exact simulation of the dynamics of the 49-atom system is no longer possible, we use the fact that two neighboring atoms cannot be excited due to the Rydberg blockade to truncate the Hilbert space from $2^{49}$ to $\sim 2^{30}$ states. We have checked with systems of up to 25 atoms, that the truncation gives the same results as an exact calculation. We solve the time-dependent Schr\"odinger equation using a split-step approach. Again, we experimentally find a deviation with respect to the spin-1/2 model for $B= 6.9$~G, while at lower $B$ the agreement is much better. We have thus identified the conditions where our system can be used as a quantum simulator of anisotropic spin-1/2 Ising model. 

In conclusion, we have explored the mapping on spin-1/2 models of interacting multilevel Rydberg atoms by taking into account the underlying details of the atomic structure in the presence of electric and magnetic fields. We searched for conditions under which the interaction between two Rydberg atoms can be faithfully described by a single potential curve. We found that this approximation can be sensitive to electric fields, thus extending previous studies on the breakdown of the Rydberg blockade~\cite{Walker2005,Reinhard2007,Walker2008,Pohl2009,Vermersch2015,Derevianko2015}, and searched numerically for an optimal region of parameters. Then, using atomic arrays of increasing size, from a pair of atoms to a $7\times7$ array, we confirmed that their dynamics under a quench is accurately reproduced by a spin-1/2 model with anisotropic Ising interaction. This work opens exciting prospects for harnessing the rich interaction spectrum of Rydberg atoms, for the engineering of various spin Hamiltonians---Ising, spin-exchange, or XXZ---as also proposed for polar molecules~\cite{Peter2012}. These insights could also help improving the control of interactions in Rydberg dressing experiments using $nP_{3/2}$ states~\cite{Biedermann2016}, as well as for Rydberg slow light polaritons with $nD$ states \cite{Hofferberth2015}. 

\begin{acknowledgements}
	This  work  benefited  from  financial  support  by  the  EU  [H2020 FET-PROACT  Project RySQ],  by  the  PALM  Labex (projects  QUANTICA and XYLOS),  and  by the R\'egion \^Ile-de-France in the framework of DIM Nano-K. 
\end{acknowledgements}

$^{\ast}$ S. de L. and S. W. contributed equally to this work.


\begin{thebibliography}{34}
\bibitem [{foo()}]{footnote_fluctuation}%
\BibitemOpen
\href@noop {} {}\bibinfo {note} {We include the residual shot-to-shot
	fluctuations of the interatomic distance by averaging the result of the
	Schr\"odinger equation over 200 realizations of the random positions of the
	atoms in their tweezers}\BibitemShut {NoStop}%
\end{thebibliography}
\end{document}